\documentclass[conference]{IEEEtran}

\usepackage[hidelinks]{hyperref}
\hypersetup{
  colorlinks   = true, 
  urlcolor     = blue, 
  linkcolor    = blue, 
  citecolor   = blue 
}
\usepackage[bottom]{footmisc}
\usepackage{graphicx}
\usepackage{fixltx2e}
\usepackage{stackengine}
\newcommand\textsub[1]{\stackengine{-.5ex}{}{\scriptsize#1}{O}{l}{F}{F}{L}}
\usepackage{subcaption}
\usepackage{cite}
\ifCLASSINFOpdf
\else
\fi
\begin{document}
%
\title{VeBPF Many-Core Architecture for Network Functions in FPGA-based SmartNICs and IoT}

\author{Zaid Tahir\textsuperscript{a}\qquad Ahmed Sanaullah\textsuperscript{b}\qquad Sahan Bandara\textsuperscript{a}\qquad Ulrich Drepper\textsuperscript{b}\qquad Martin Herbordt\textsuperscript{a}\\\textsuperscript{a}CAAD Lab,\ Electrical\ and\ Computer\ Engineering, Boston\ University,\ USA.\  -\  \textsuperscript{b}Red\ Hat\ Inc.\\zaidt@bu.edu\quad asanaull@redhat.com\quad sahanb@bu.edu\quad drepper@redhat.com\quad herbordt@bu.edu}


\maketitle

\begin{abstract}
 FPGA-based SmartNICs and IoT devices integrated with soft-processors for executing network functions have been introduced to overcome hardware-reconfigurability limitations in DPUs (Data Processing Units) and MCUs (Microcontroller Units), respectively. However, existing FPGA-based SmartNICs and IoT devices lack a highly configurable many-core architecture that specializes in network packet processing.
 
 This work introduces a resource-optimized highly configurable VeBPF (Verilog eBPF) many-core architecture built upon VeBPF CPU cores that we have developed for specialized network packet processing in FPGAs. These VeBPF cores are eBPF ISA compliant and have been developed in Verilog HDL for easy integration with existing FPGA IP blocks/subsystems. The VeBPF many-core architecture executes multiple eBPF rules on multiple VeBPF cores in-parallel for low-latency network packet processing. Due to the highly configurable hardware design of this VeBPF many-core architecture, any number of VeBPF cores can be instantiated by assigning a parameter \textit{N\textsub{VeBPF}} in the Verilog code of the VeBPF many-core architecture and any number of eBPF rules can be uploaded, with FPGA resources as the only constraint. The proposed VeBPF many-core architecture has been designed to process eBPF rules faster if \textit{N\textsub{VeBPF}} is increased and the eBPF rules can be dynamically changed during run-time without requiring new bitstreams. It uses various hardware and computer architecture optimizations to support its implementation on low-end FPGAs-based IoT devices along with high-end FPGA-based SmartNICs, for network packet processing. We have also developed automatic-testing and simulation frameworks for the proposed VeBPF many-core architecture, using the latest open-source tools like Python and Cocotb. We have released the Verilog HDL code for VeBPF core development, VeBPF many-core architecture, C software libraries for RISC-V control of m-plane (management-plane) of the VeBPF many-core architecture and the simulators as an open-source contribution for further advancement of FPGAs in many-core architectures, eBPF, SmartNICs, IoT, cybersecurity and communication.            
 
\end{abstract}

\IEEEpeerreviewmaketitle

\section{Introduction and background}

As is well-known, advances in process technology have run up against limitations in Dennard scaling and Moore's Law resulting in fundamental changes to CPU architecture, the most obvious being the emergence of multicore. Other fundamental shifts in computing, such as to data-centers/clouds and edge-based IoT devices have exposed new limitations in CPU architectures, especially when cost and power are considered. Money, energy and time is lost with every cycle of cloud host CPUs spent on network communication and other tasks that are unrelated to the user applications.


In order to decouple host CPUs from the computational loads of network packet processing, SmartNICs (Smart Network Interface Cards) \cite{forencich2020corundum} have been introduced that perform network functions. Since SmartNICs need high computational capability, throughput, and energy efficiency, many-core processors have also been integrated into the latest SmartNICs such as DPUs (Data Processing Units) \cite{DPU} or FPGA-based SmartNICs \cite{lin2020panic}. 

DPU-based SmartNICs suffer from the limitation that the hardware is fixed and if new hardware features are required, e.g., due to upgrades in protocols and interfaces, the existing DPU may need to be replaced, or, at least, lose relative performance.
FPGA-based SmartNICs have long filled a niche in this space as they provide high-throughput communication while having reconfigurable hardware \cite{Xiong18a,Xiong19,Guo22,Guo22a}. Often this configurability is used, at least partially, to implement dedicated soft processors \cite{lin2020panic,kamaleldin2020towards,Guo23}. The issue addressed here is the design of these processors: many-core architectures available for FPGAs are mostly based on homogeneous general purpose processors such as RISC-V \cite{gray2016grvi}, \cite{kamaleldin2020towards}, or, if they are heterogeneous many-core architectures \cite{kurth2017hero}, they still involve various general purpose processors not specialized for network packet processing.

For network packet processing eBPF ISA (Instruction Set Architecture) \cite{BPF} provides specialized instructions for network functions, which is one of the reasons eBPF compilers and tool-chains are native to UNIX-like operating systems with eBPF bytecode executed in the kernel space. Many technology companies have included eBPF into their software stacks \cite{eBPF}. Due to these advantages eBPF ISA compliant soft-processors have been developed like \cite{brunella2022hxdp,hBPF}. Both, however are single processor solutions. \cite{brunella2022hxdp} uses a custom compiler to convert native eBPF instructions to special VLIW instructions; this introduces issues of maintaining correct versions of compilers and drivers \cite{Bandara22,Bandara24,Bandara24c}. \cite{hBPF} is written in Migen \cite{migen}, which may be difficult to integrate 
with IP blocks written in HDLs. 

In the existing FPGA-based SmartNICs such as \cite{forencich2020corundum,lin2020panic,firestone2018azure,zhao2020achieving}, the network packet processing frameworks are aimed at high-end FPGAs for cloud-based deployments. This leaves a void for network packet processing frameworks for low-end FPGAs essential to IoT devices \cite{yang2021fpga,catarinucci2015iot}. These FPGA-based IoT devices have easy access to fast wireless networks due to the maturity of 5G infrastructure. The communication loads on these FPGA-based IoT devices have increased tremendously, especially with AI integrated in many applications. Such high connectivity also introduces threats of malicious cyberattacks \cite{wood2002denial}. Hence, low-end FPGAs used as IoT devices also need a network packet processing framework that takes the network packet processing load off the main soft-processor. The problem is that these low-end FPGA-based IoT devices lack such a resource-optimized network packet processing framework.

Due to the lack of many-core architectures specialized in network packet processing and in order to offload network communication processing loads for both high-end FPGA-based SmartNICs and low-end FPGA-based IoT devices, this paper presents a highly configurable and resource-optimized VeBPF many-core architecture. The PE (Processing Element) of the proposed many-core architecture, the VeBPF CPU core, has been developed to specialize in network packet processing.

The contributions of this work are summarized as follows:
\begin{itemize}
    \item The PE of the proposed many-core architecture, the VeBPF core, has been developed to be eBPF ISA \cite{BPF} compliant, making it specialized for network packet processing. It is implemented in Verilog HDL, ensuring portability and easy integration with existing IP blocks.
    \item We have developed various design optimizations for low-latency network packet processing, e.g., single clock cycle reprogramming of the VeBPF core, making it possible to switch between multiple eBPF rules by changing the VeBPF core PC externally in just a single clock cycle.
    \item Due to the flexible and optimized hardware design of the VeBPF many-core architecture, any number of VeBPF cores can be instantiated and any number of eBPF rules can be uploaded. 
    \item The scheduling and arbitration logic in the proposed VeBPF many-core architecture has been designed in such a way that increasing the number of VeBPF cores 
    leads to faster processing of the eBPF rules on the network packets.
    \item We have designed the shared data and control buses of the VeBPF many-core designs to minimize resource usage. 
    Resource-intensive communication modules such as NOCs and reconfigurable match action pipelines (RMTs) were intentionally avoided.
    \item Unlike existing network packet processing frameworks in FPGA-based SmartNICs \cite{forencich2020corundum,lin2020panic}, where the full network packets are held in temporary registers till processing on them is completed, the VeBPF many-core architecture has been designed such that the full network packets are written to memory as soon as they arrive.
    \item We also developed an automatic testing framework for the developing and upgrading the VeBPF CPU core, and a complete simulation framework for the VeBPF many-core designs.
\end{itemize}

We have released the Verilog HDL code for VeBPF core\footnote{\href{https://github.com/zaidtahirbutt/VeBPF}{https://github.com/zaidtahirbutt/VeBPF}} development and the VeBPF many-core design.\footnote{\href{https://github.com/zaidtahirbutt/VebpfManyCore}{https://github.com/zaidtahirbutt/VebpfManyCore}} This includes the C libraries for RISC-V control of the management-plane, as well as the simulation and testing frameworks. 

In the coming sections we elaborate on the computer architecture design details of the VeBPF core development and the overall many-core hardware architecture. We also present a firewall application implemented using VeBPF on a FPGA-based IoT device. We compare resource usage of the VeBPF cores with similar PEs. 

\section{VeBPF Many-Core Architecture}

This section presents a detailed overview of the hardware architecture of the proposed VeBPF many-core architecture and its building blocks, starting from the many-core PE (the VeBPF CPU core) and rest of the modules that make up this VeBPF many-core architecture.

\subsection{VeBPF CPU Core}

One of the major goals of the proposed VeBPF many-core architecture is to ensure that the PE, the basic compute unit of the many-core architecture, is specialized for network packet processing and the VeBPF many-core architecture is readily usable without requiring custom drivers and compilers. It is due to these reasons that we developed the VeBPF CPU core, which is the PE for our VeBPF many-core architecture, to be eBPF ISA \cite{BPF} compliant since eBPF is native to UNIX-like OS and is widely used by technology companies in their software stacks \cite{eBPF}. 

We deliberately chose Verilog as the HDL language for the development of the VeBPF CPU core, after doing research and development on other higher-level HDL languages like Migen \cite{migen}, \cite{hBPF}, we found that these higher-level HDL languages aren’t readily portable with other lower-level HDL languages like Verilog, VHDL and System Verilog. Hence developing the VeBPF CPU core in Verilog HDL would make our VeBPF many-core architecture quite portable.  

\begin{figure}[h]
\centering
\includegraphics[width=\columnwidth]{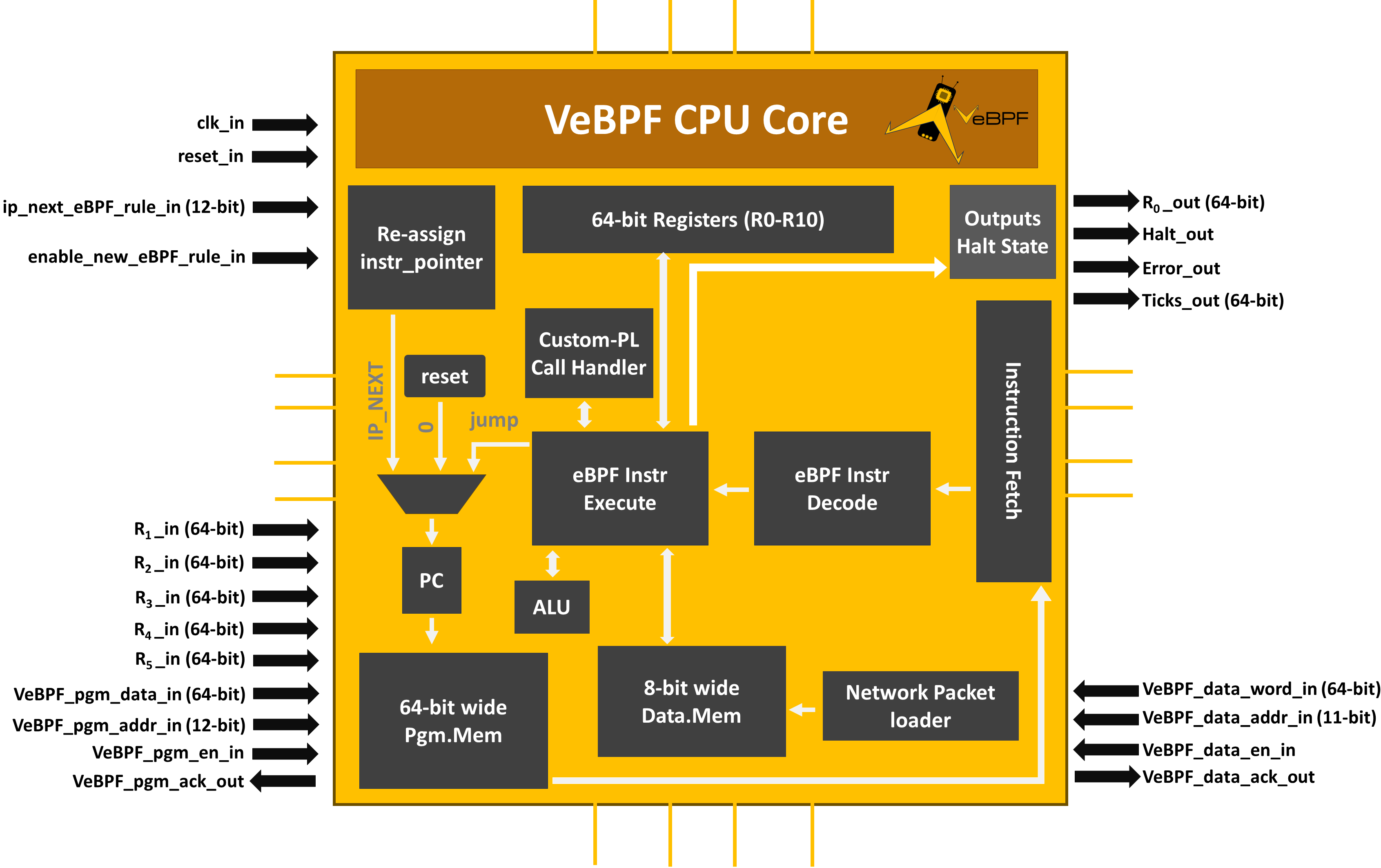}
\caption{VeBPF CPU core computer architecture overview.}
\label{fig1:VeBPF_cpu_core}
\end{figure}

\subsubsection{VeBPF CPU core computer architecture}
The VeBPF CPU core is depicted in Fig.~\ref{fig1:VeBPF_cpu_core}, has a Harvard architecture as seen from the separate program and data memory blocks. The data memory is 8-bits wide and its depth is adjustable. We normally select the depth of the data memory to be equal to that of the of the maximum length of the network packet headers instead of setting it to full network packet length because the VeBPF core only needs to process the headers of the network packets as is usually done in eBPF programs. The program memory is 64-bits wide and its depth is also adjustable. The VeBPF core has 11 64-bit registers (\textit{R0 – R10}). The registers \textit{R1 – R5} are used as input registers and they are not cleared upon \textit{reset} given to the VeBPF core. The register \textit{R0} is used as the output register.

As soon as the \textit{reset\_in} signal is set \textit{LOW}, the VeBPF core starts executing the eBPF instructions that are read from the program memory based on the value of \textit{PC} and sent to the \textit{Instruction Fetch} module, which then get decoded and executed by their respective blocks as shown in Fig.~\ref{fig1:VeBPF_cpu_core}. As soon as the eBPF \textit{exit} instruction is processed and the eBPF program is finished, the \textit{Halt\_out} output port becomes \textit{HIGH} with the output result available at \textit{R0} register. If any error had occurred during processing, the \textit{Error\_out} signal becomes \textit{HIGH}. The total clock ticks taken to run the eBPF program are also available at \textit{Ticks\_out} output port. The eBPF \textit{call} instruction in Linux Kernel is used to call \textit{Helper\_Functions} while in case of the VeBPF core, the \textit{call} eBPF instruction is directed to the \textit{Custom-PL Call Handler} block where users can implement application specific custom hardware accelerators.

\subsubsection{Single clock cycle eBPF rule switching}
We have developed an important functional optimization to the computer architecture of the VeBPF core where the instruction pointer that reads the eBPF program instruction from the program memory, is settable from outside the VeBPF core while the VeBPF core is in reset state, for the purpose of reprogramming the VeBPF CPU core to a different eBPF rule within a single clock cycle. The block labelled as \textit{Re-assign instr\_pointer} depicted in Fig.~\ref{fig1:VeBPF_cpu_core} is in charge of doing that. The benefit of this optimization is that the VeBPF core is reprogrammable to a different eBPF rule with just a change of its instruction pointer to the value where the particular eBPF rule is located in the program memory of the VeBPF core in a single clock cycle. This optimization saves tremendous amount of clock cycles versus what it would have taken if, for each eBPF rule, that rule would have had to be uploaded in the program memory of the VeBPF core first before its execution. It is pertinent to mention here that the VeBPF many-core multi-rule program loader module Fig.~\ref{fig5:VeBPF_programloader} uploads all the eBPF rules to the program memories of all VeBPF cores so that all the VeBPF cores are reprogrammable to any eBPF rule with just a change in their instruction pointers in a single clock cycle by the VeBPF many-core multi-rule scheduler module Fig.~\ref{fig6:VeBPF_sceduler}. 

\subsubsection{VeBPF CPU core resource usage}
Table-\ref{tab1:VeBPF_resources} shows the FPGA resource utilization of a single VeBPF core compared with PE of a RISC-V-based many-core architecture \cite{kamaleldin2020towards} and a eBPF ISA compliant core that needs custom compilers and drivers \cite{brunella2022hxdp}. Our VeBPF core requires significantly less FPGA resources as seen in Table-\ref{tab1:VeBPF_resources} which falls in-line with the goal of optimized computer architecture design of the PE of our VeBPF many-core architecture so that we are able to target both low-end FPGAs for IoT deployments and high-end FPGAs for SmartNIC deployments.

\begin{table}
    \centering
    \caption{FPGA Resource Utilization for Various PE Cores}
    \begin{tabular}{|c|c|c|c|} \hline 
 Single PE Core & LUTs & FFs & BRAMs \\ \hline  VeBPF core&  3500& 1600& 1.5\\ \hline RISC-V PE\cite{kamaleldin2020towards}&  7878& 1944& 20\\ \hline
  SEPHIROT\cite{brunella2022hxdp}&  27000& 4000& -\\ \hline
    \end{tabular}
    \label{tab1:VeBPF_resources}
\end{table}

\subsection{VeBPF Many-Core Architecture Overview}
\begin{figure}[h]
\centering
\includegraphics[width=\columnwidth]{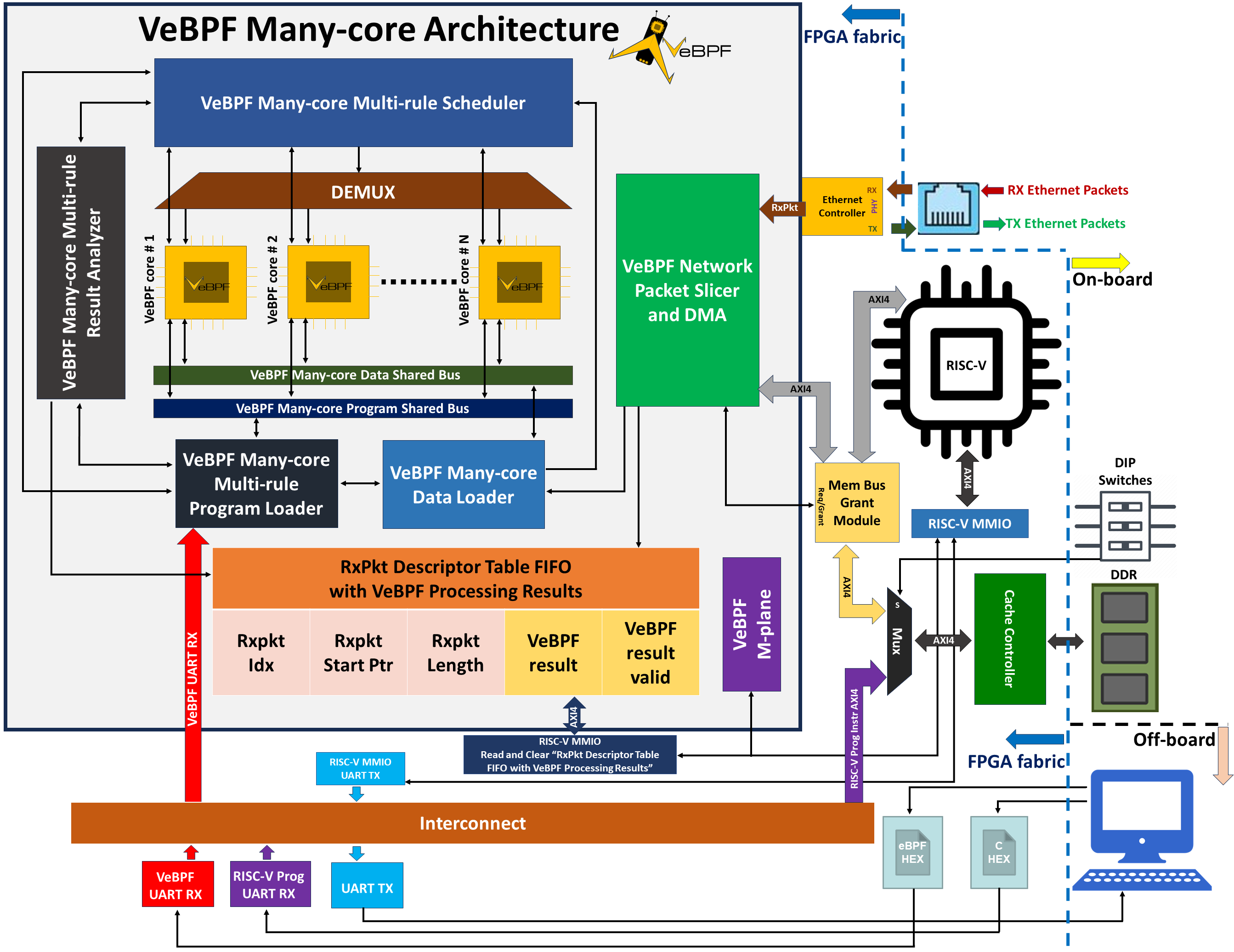}
\caption{Overview of the VeBPF many-core architecture.}
\label{fig2:VeBPF_arch}
\end{figure}

The VeBPF many-core architecture depicted in Fig.~\ref{fig2:VeBPF_arch} features VeBPF cores as the basic PE of this many-core architecture along with the main building block modules. The PEs of this VeBPF many-core architecture are accessed through different shared data buses for writing network packet data as PE data-memory and eBPF rules as PE program-memory. The PEs are also controlled and monitored through different control buses for various control operations like re-programming a certain PE, monitoring if any PE has finished processing by checking \textit{Halt\_out} output port and if it is HIGH then reading the result from its \textit{R0} register along with \textit{Error\_out} and \textit{Ticks\_out} output ports, executing a certain eBPF rule on a PE selected through arbitration, etc. All of these shared data and control buses are operated by the different building blocks of this VeBPF many-core architecture that we refer to as modules. 

All the building blocks of the proposed VeBPF many-core architecture including the PEs and modules, are highly flexible and can be adjusted for various target deployments, e.g.,  if a high performance deployment is needed like a SmartNIC, then the modules can be configured for that, similarly they can also be configured for low-resource deployment like IoT.

The proposed VeBPF many-core architecture sits between the network subsystem and the memory subsystem as shown in Fig.~\ref{fig2:VeBPF_arch}. The RISC-V soft-processor controls the m-plane, i.e., the CSRs (control status registers) of the VeBPF many-core architecture. The complex interactions of RISC-V with the VeBPF many-core architecture on the m-plane are available in the C libraries we provide with the rest of the open-source code for this VeBPF many-core architecture.   

The VeBPF many-core architecture is connected with the UART subsystem as well. You can see in Fig.~\ref{fig2:VeBPF_arch} that the VeBPF many-core architecture and the RISC-V soft-processor are sharing the same DDR memory, it is because the target deployment for the VeBPF many-core architecture shown in this Fig.~\ref{fig2:VeBPF_arch} is for the FPGA-based IoT device deployment for which we have performed various experiments for this paper. If a different deployment was targeted like a FPGA-based SmartNIC, the VeBPF many-core architecture would have had its own separate memory to write packets to, whereas the RISC-V would have had its own separate memory. The aim of showing this particular shared memory configuration is to show the flexibility of our VeBPF many-core architecture and its ability to provide native eBPF rule-based network packet processing functionality to a resource limited FPGA-based IoT device as well as for high-end FPGA-based SmartNICs.

Before going into the details of the main modules of the VeBPF many-core architecture, a summary of the main steps required to activate the VeBPF many-core architecture is listed below:

1) The number of VeBPF cores required just need to be specified in the \textit{N\textsub{VeBPF}} parameter before compilation of the FPGA bitstream. More VeBPF cores would result in faster processing of the eBPF rules;

2) The eBPF rules are uploaded from the host PC through the UART subsystem (\textit{VeBPF UART RX}) as shown in Fig.~\ref{fig2:VeBPF_arch}. Any number of eBPF rules can be uploaded and the eBPF rule-set can be changed dynamically during run-time.

3) The RISC-V program instructions for controlling the m-plane of the VeBPF many-core architecture are uploaded through the UART subsystem as well (\textit{RISC-V Prog UART RX}) as shown in Fig.~\ref{fig2:VeBPF_arch}. After RISC-V is activated, as a part of the m-plane functionality, it allocates the starting memory address and the total memory available for the network packets to their respective CSRs in the VeBPF many-core architecture through the MMIO bus connection. This step arms the VeBPF many-core architecture to start receiving network packets from the network subsystem;

Rest of the steps of operations are mentioned in the modules definitions of the VeBPF many-core architecture below.

\subsection{VeBPF Network Packet Slicer and DMA}

\begin{figure}[h]
\centering
\includegraphics[width=\columnwidth]{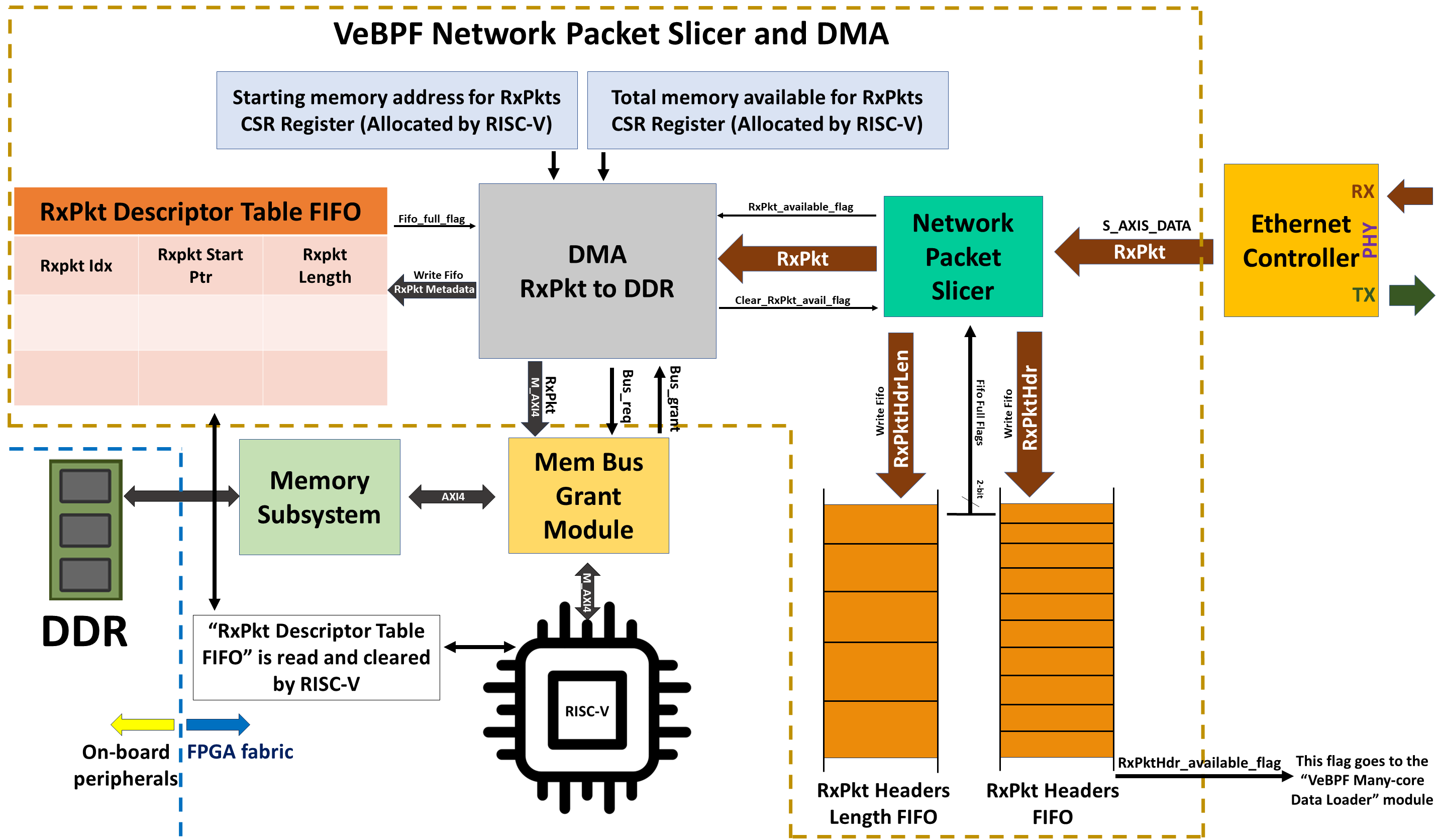}
\caption{VeBPF Network Packet Slicer and DMA Module}
\label{fig3:VeBPF_Packet_Slicer_DMA}
\end{figure}

The VeBPF network packet slice and DMA (Direct Memory Access) module is shown in detail in Fig.~\ref{fig3:VeBPF_Packet_Slicer_DMA}. This module is responsible for receiving the network packets from the network subsystem (\textit{Ethernet Controller} block) through an AXI-stream port. Unlike other network packet processing frameworks  \cite{lin2020panic}, \cite{brunella2022hxdp} that have to hold the network packets till processing on those packets is completed, the \textit{Network Packet Slicer} slices the header of the network packet (\textit{RxPkt}) and copies the network packet header and its header length in two FIFOs while handing-off the full network packet to the \textit{DMA RxPkt to DDR} block. It is important to note that the header length varies according to the type of network packet received and users can set a custom header length as well. 

The \textit{DMA RxPkt to DDR} block checks the CSRs for the starting memory address and total memory available for the network packets. The \textit{DMA RxPkt to DDR} block writes the \textit{RxPkt} to memory after acquiring the grant for the memory bus from the \textit{Mem Bus Grant Module} block and writes the corresponding \textit{RxPkt} metadata to the \textit{RxPkt Descriptor Table FIFO} and updates its own registers that keep track of of the currently available memory and the current memory address for the next \textit{RxPkt} to write to. The RISC-V using the m-plane is able to read the \textit{RxPkt} metadata from the \textit{RxPkt Descriptor Table FIFO}. The RISC-V can then access the \textit{RxPkts} from memory if needed, using that metadata. The RISC-V through the m-plane then clears the \textit{RxPkt} metadata \textit{RxPkt Descriptor Table FIFO} entries by incrementing the relevant FIFO read pointers, which signals to the \textit{DMA RxPkt to DDR} block to increment its currently available memory. The VeBPF Network Packet Slicer and DMA module sends a HIGH \textit{RxPktHdr\_available\_flag} signal to the VeBPF many-core data loader module (Fig.~\ref{fig4:VeBPF_dataloader}) as soon as a \textit{RxPkt} header becomes available in the \textit{RxPkt Headers FIFO}. 

\subsection{VeBPF Many-core Data Loader}

\begin{figure}[h]
\centering
\includegraphics[width=\columnwidth]{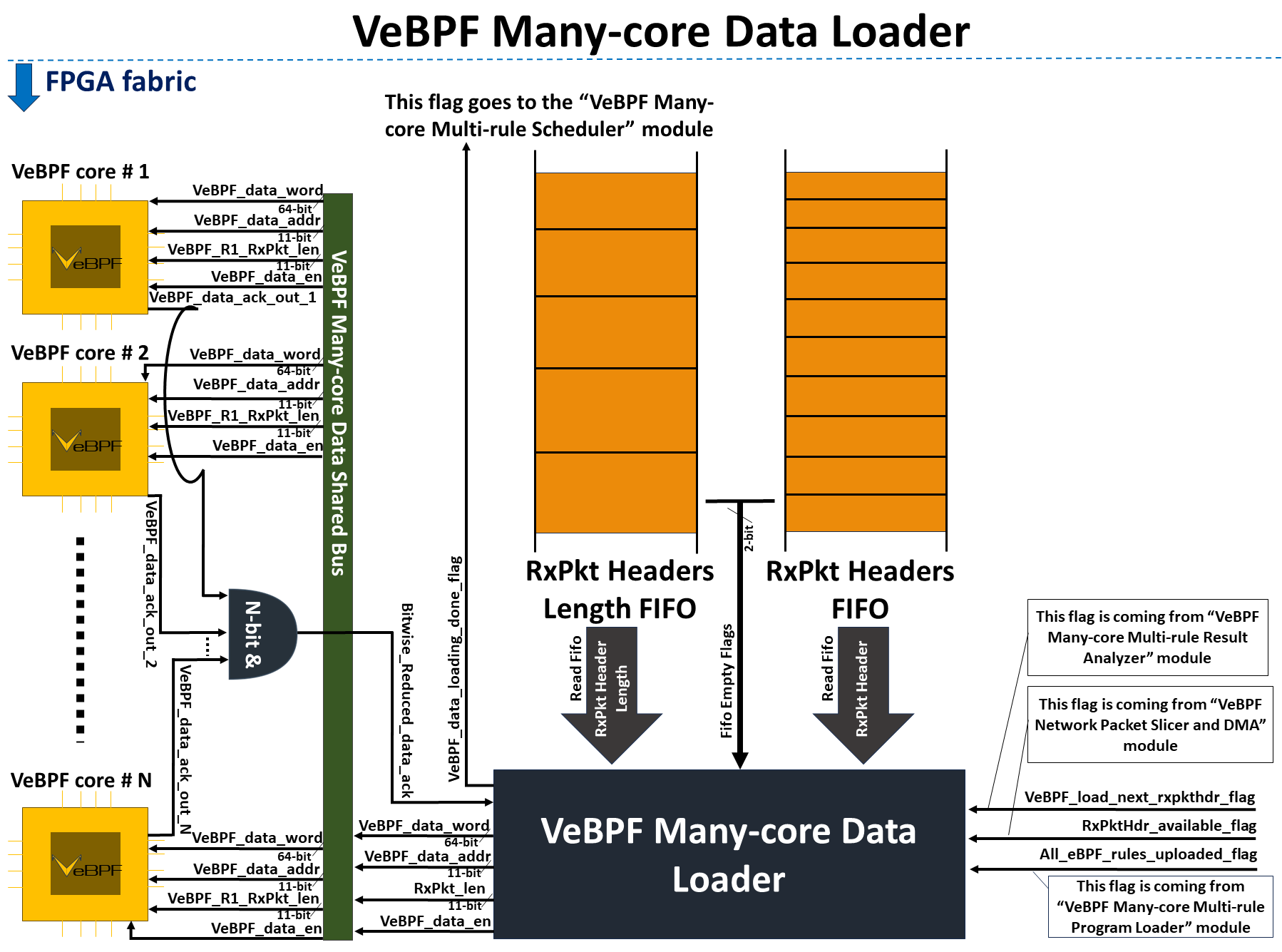}
\caption{VeBPF Many-core Data Loader Module}
\label{fig4:VeBPF_dataloader}
\end{figure}

The VeBPF many-core data loader module reads the \textit{RxPkt} header and its length from the FIFOs, that were written to by the VeBPF network packet slicer and DMA module, as soon as it receives a HIGH \textit{RxPktHdr\_available\_flag}. The VeBPF many-core data loader loads the \textit{RxPkt} header in all the VeBPF cores through the \textit{VeBPF Many-core Data Shared Bus} as shown in Fig.~\ref{fig4:VeBPF_dataloader}. 

For each 64-bit data word of the \textit{RxPkt} header that is written to all the VeBPF cores, the \textit{VeBPF Many-core Data Loader} block waits for acknowledgements (ACKs) from all the VeBPF cores using a reduction bit-wise AND operation on the ACK signals from all VeBPF cores, before the \textit{VeBPF Many-core Data Loader} block writes the next 64-bit data word of the \textit{RxPkt} header till the full header is written to all the VeBPF cores data memories. The reason for writing the same \textit{RxPkt} header in all the VeBPF cores is that multiple eBPF rules can be executed in-parallel on different VeBPF cores on the same \textit{RxPkt} header so that a valid result can be obtained in minimal time, in order to keep the processing latency to a minimum. 

After the VeBPF many-core data loader module uploads a \textit{RxPkt} header to all the VeBPF cores, it sends a HIGH \textit{VeBPF\_data\_loading\_done\_flag} signal to VeBPF many-core multi-rule scheduler module, indicating that there is a \textit{RxPkt} header available for executing eBPF rules on. 

\subsection{VeBPF Many-core Multi-rule Program Loader}

The VeBPF many-core multi-rule program loader module is what gives the VeBPF many-core architecture the flexibility to upload eBPF rules of varying lengths and the dynamicity to change the eBPF rule-set during run-time without requiring a new FPGA bitstream. 

\begin{figure}[h]
\centering
\includegraphics[width=\columnwidth]{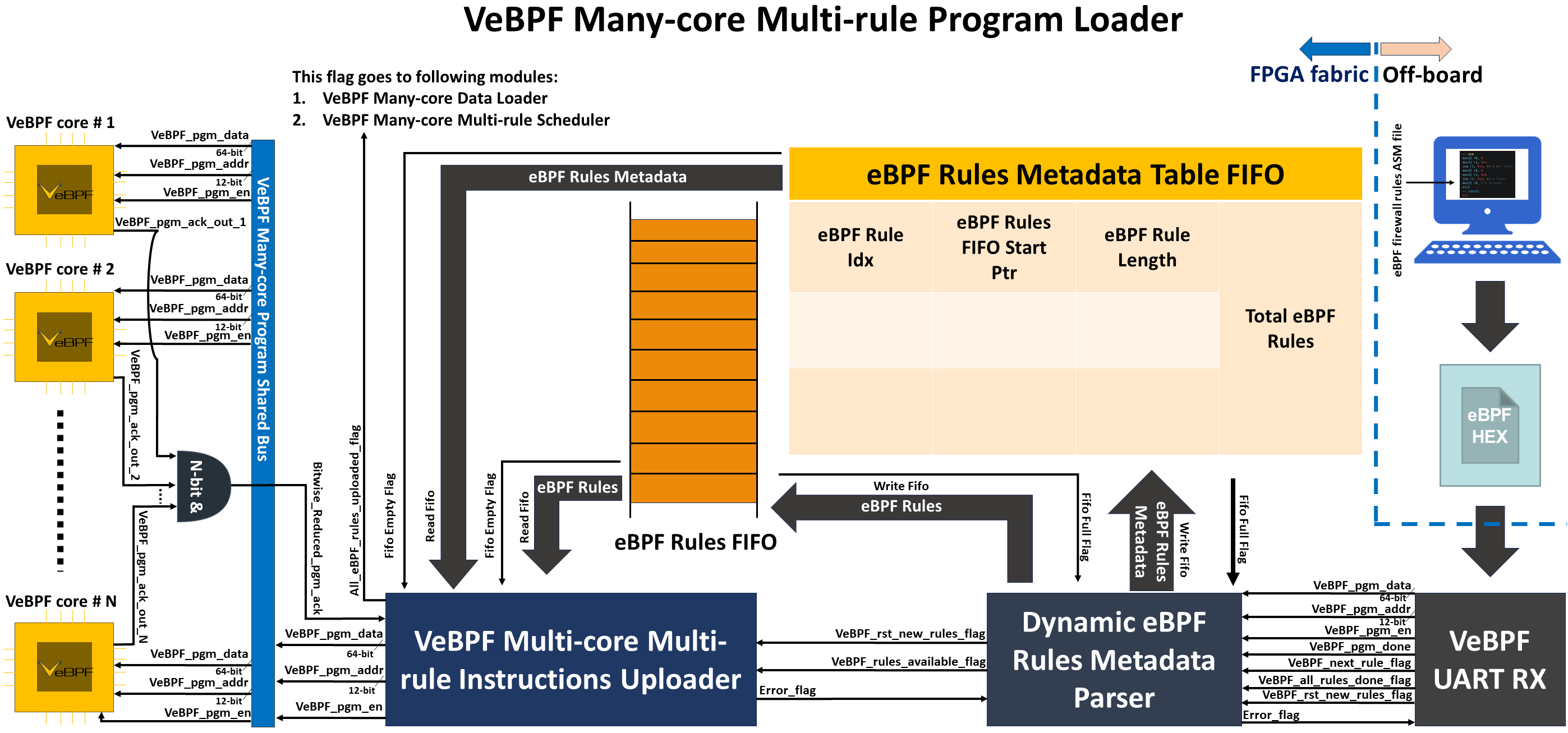}
\caption{VeBPF many-core multi-rule program loader module}
\label{fig5:VeBPF_programloader}
\end{figure}

The VeBPF many-core multi-rule program loader module is depicted in Fig.~\ref{fig5:VeBPF_programloader}. The eBPF rules uploaded enter from the UART subsystem (\textit{VeBPF UART RX}) and since the eBPF rules can be of any length, the \textit{Dynamic eBPF Rules Metadata Parser} block stores these eBPF rules in a \textit{eBPF Rules FIFO} along with the metadata of those corresponding eBPF rules in the \textit{eBPF Rules Metadata Table FIFO}. 

Once all eBPF rules have been uploaded by the user, the \textit{VeBPF Multi-core Multi-rule Instruction Uploader} block reads these eBPF rules and their metadata from the corresponding FIFOs and these eBPF rules are uploaded as program memory to all VeBPF cores through a shared \textit{VeBPF Many-core Program Shared Bus}. The ACKs from all the VeBPF cores are bit-wise reduced AND-ed together after each 64-bit eBPF instruction program word is written to every VeBPF core, and this process repeats till all eBPF rules are present in every VeBPF core program memory. The reason for all VeBPF cores having all the eBPF rules in their program memory is that the VeBPF many-core multi-rule scheduler module Fig.~\ref{fig6:VeBPF_sceduler} can switch to any eBPF rule on any VeBPF core in a single clock cycle by just changing the instruction pointer of that VeBPF core. 

If a new eBPF rule-set is required, a HIGH \textit{VeBPF\_rst\_new\_rules\_flag} signal is sent (Fig.~\ref{fig5:VeBPF_programloader}) and the whole eBPF rules uploading process is repeated again. Once all eBPF rules have been uploaded to all VeBPF cores, the VeBPF many-core multi-rule module sends a HIGH \textit{All\_eBPF\_rules\_uploaded\_flag} signal to the VeBPF many-core data loader module and the VeBPF many-core multi-rule scheduler module and the flag \textit{All\_eBPF\_rules\_uploaded\_flag} is kept HIGH unless a new eBPF rule-set is being uploaded. 

\subsection{VeBPF Many-core Multi-rule Scheduler}

\begin{figure*}
    \centering
    \centerline{\includegraphics[width=1\textwidth]{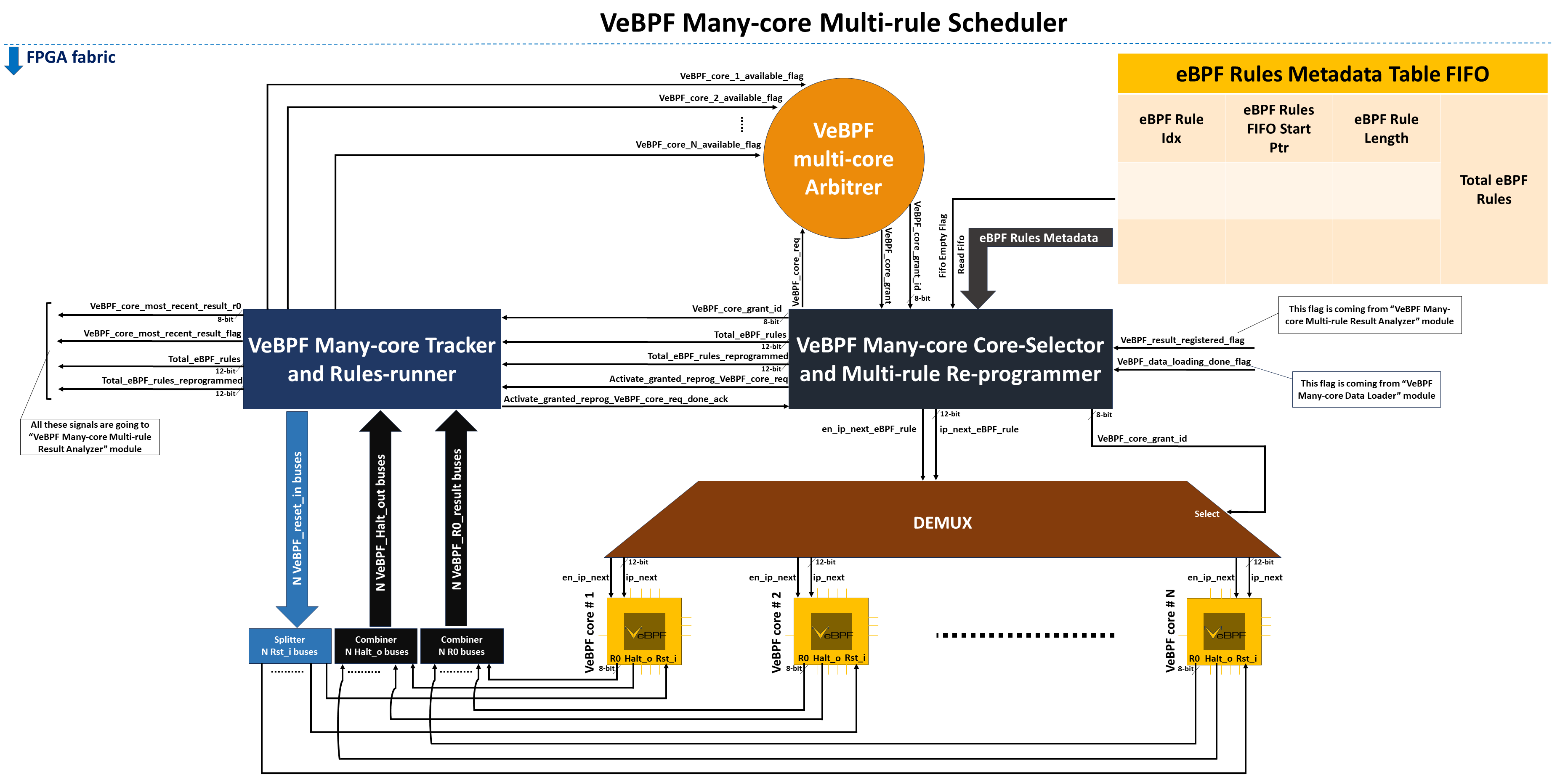}}
    \caption{VeBPF many-core multi-rule scheduler module}
    \label{fig6:VeBPF_sceduler}
\end{figure*}

This VeBPF many-core multi-rule scheduler module depicted in Fig.~\ref{fig6:VeBPF_sceduler} is one of the most complex hardware logic in the VeBPF many-core architecture that really makes this architecture highly parallelized, resource-efficient and optimized for low latency multi-rule eBPF network packet processing. 

The VeBPF many-core multi-rule scheduler module waits for HIGH flags from VeBPF many-core data loader module (\textit{VeBPF\_data\_loading\_done\_flag}) and VeBPF many-core multi-rule program loader module (\textit{All\_eBPF\_rules\_uploaded\_flag}), HIGH values of these flags indicate that a \textit{RxPkt} header is available for processing in all VeBPF cores and all eBPF rules have been uploaded in all VeBPF cores respectively. 

The VeBPF many-core multi-rule scheduler then checks the \textit{VeBPF multi-core Arbitrer} block to see if any VeBPF core is idle, and as soon as the arbitrer provides an idle VeBPF core to the \textit{VeBPF Many-core Core-Selector and Multi-rule Re-programmer} block, it reprograms that idle VeBPF core to the current eBPF rule, that hasn't been executed yet, using the eBPF rule metadata from the \textit{eBPF Rules Metadata Table FIFO} that was filled by the VeBPF many-core multi-rule program loader module. The idle VeBPF core is reprogrammed to this selected eBPF rule in a single clock cycle by first selecting that VeBPF core through a DEMUX using the grant id given by the \textit{VeBPF multi-core Arbitrer} as the select line of the DEMUX as shown in Fig.~\ref{fig6:VeBPF_sceduler}. 

After the idle VeBPF core has been accessed through the DEMUX, it is reprogrammed to the selected eBPF rule in a single clock cycle by changing its instruction pointer externally to the location of the selected eBPF rule. After the \textit{VeBPF Many-core Core-Selector and Multi-rule Re-programmer} block reprograms the idle VeBPF core, it transfers the grant id and the relevant eBPF rules information like total eBPF rules already reprogrammed, to the \textit{VeBPF Many-core Tracker and Rules-runner} block which basically activates/runs the reprogrammed idle VeBPF core and keeps a track of it until it reaches a \textit{halt} state. 

As soon as the VeBPF cores reach \textit{halt} states, the \textit{VeBPF Many-core Tracker and Rules-runner} block sends the VeBPF \textit{R0} result information and the eBPF rules information, like total eBPF  rules already reprogrammed, to the VeBPF many-core multi-rule result analyzer module (Fig.\ref{fig7:VeBPF_result_analyzer}). While the \textit{VeBPF Many-core Tracker and Rules-runner} block is running and keeping track of all the VeBPF cores in parallel and forwarding their results to the VeBPF many-core multi-rule result analyzer module as soon as the results are received, the \textit{VeBPF Many-core Core-Selector and Multi-rule Re-programmer} block is incrementing the eBPF rule index as soon as it hands off a reprogrammed VeBPF core to the \textit{VeBPF Many-core Tracker and Rules-runner} block, and then \textit{VeBPF Many-core Core-Selector and Multi-rule Re-programmer} block waits for the \textit{VeBPF multi-core Arbitrer} block to give it the next idle VeBPF core so that the next eBPF rule can be reprogrammed on the granted idle VeBPF core and handed off to the \textit{VeBPF Many-core Tracker and Rules-runner} block. All of these hand-shakes between \textit{VeBPF Many-core Core-Selector and Multi-rule Re-programmer} and \textit{VeBPF Many-core Tracker and Rules-runner} blocks are happening in parallel till the VeBPF many-core multi-rule result analyzer module sends a valid result received flag (\textit{VeBPF\_result\_registered\_flag}).

After the VeBPF many-core multi-rule scheduler module receives the valid result flag \textit{VeBPF\_result\_registered\_flag} from the VeBPF many-core multi-rule result analyzer module, it waits for the next \textit{RxPkt} header to be uploaded by the VeBPF many-core data loader module and this whole process repeats that involves reprogramming and running eBPF rules on the multiple VeBPF cores in parallel till a valid eBPF packet processing result is received.   


\subsection{VeBPF Many-core Multi-rule Result Analyzer}
\begin{figure}[h]
\centering
\includegraphics[width=\columnwidth]{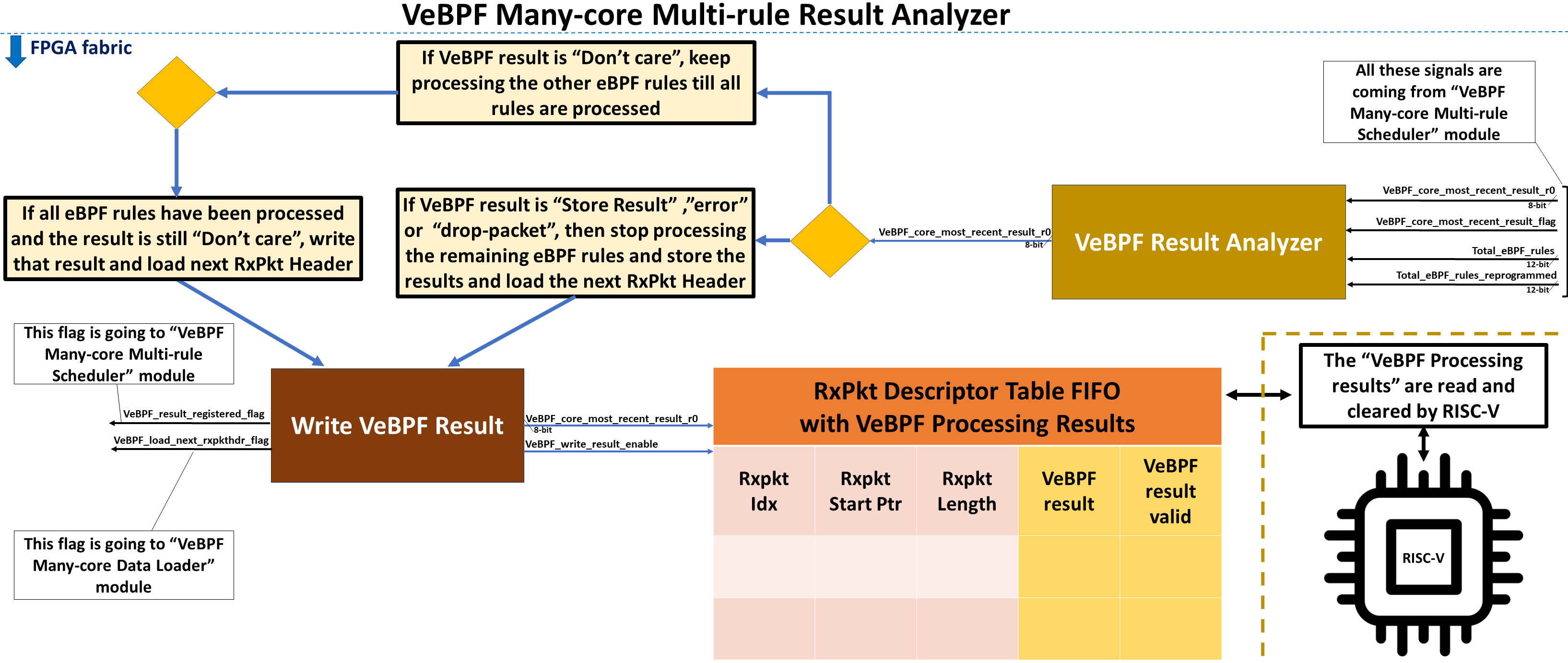}
\caption{VeBPF many-core multi-rule result analyzer module}
\label{fig7:VeBPF_result_analyzer}
\end{figure}

The VeBPF many-core multi-rule result analyzer module is depicted in Fig.~\ref{fig7:VeBPF_result_analyzer} and this module takes in inputs from the VeBPF many-core multi-rule scheduler module and these inputs include the most recent eBPF processing result \textit{VeBPF\_core\_most\_recent\_result\_r0} along with total eBPF rules \textit{Total\_eBPF\_rules} and total eBPF rules that have been run/reprogrammed currently \textit{Total\_eBPF\_rules\_reprogrammed}. 

The data flow diagram in Fig.~\ref{fig7:VeBPF_result_analyzer} tells us how VeBPF many-core multi-rule result analyzer module works. If a valid eBPF result (valid result is either "store result", "error", "drop packet") is received in the \textit{VeBPF Result Analyzer} block or if the eBPF result is "don't care" but the \textit{Total\_eBPF\_rules\_reprogrammed} is equal to \textit{Total\_eBPF\_rules}, then forward the eBPF processing result to the \textit{Write VeBPF Result} block.

The \textit{Write VeBPF Result} block appends the VeBPF result to the \textit{RxPkt Descriptor Table FIFO} which now is renamed as \textit{RxPkt Descriptor Table FIFO with VeBPF Processing Results}. The RISC-V soft-processor through the m-plane, can read the VeBPF processing results directly from this \textit{RxPkt Descriptor Table FIFO with VeBPF Processing Results} table instead of having to read the \textit{RxPkt} metadata from this table and using this metadata, read the \textit{RxPkt} from memory and processing the packet as per the eBPF rules for the same result. 

The \textit{Write VeBPF Result} block after writing the VeBPF results sends HIGH \textit{VeBPF\_result\_registered\_flag} signal to VeBPF many-core multi-rule scheduler module so it can start processing eBPF rules on the next \textit{RxPkt} header. \textit{Write VeBPF Result} block also sends a HIGH \textit{VeBPF\_load\_next\_rxpkthdr\_flag} signal to the VeBPF many-core data loader module signalling it to load the next \textit{RxPkt} header into the VeBPF cores data memory. 

As seen in the detailed description of the hardware architecture of the VeBPF many-core architecture modules, the limited space in this paper isn't enough to highlight every important detail, so the detailed figures along with looking at our open-source code for this VeBPF many-core architecture would be useful for further insights. The  C code libraries for the RISC-V control of the m-plane of the VeBPF many-core architecture are also an important part of the body of knowledge of this VeBPF many-core architecture.

\subsection{Automatic Testing \& Simulation Framework }
\begin{figure}[h]
\centering
\includegraphics[width=\columnwidth]{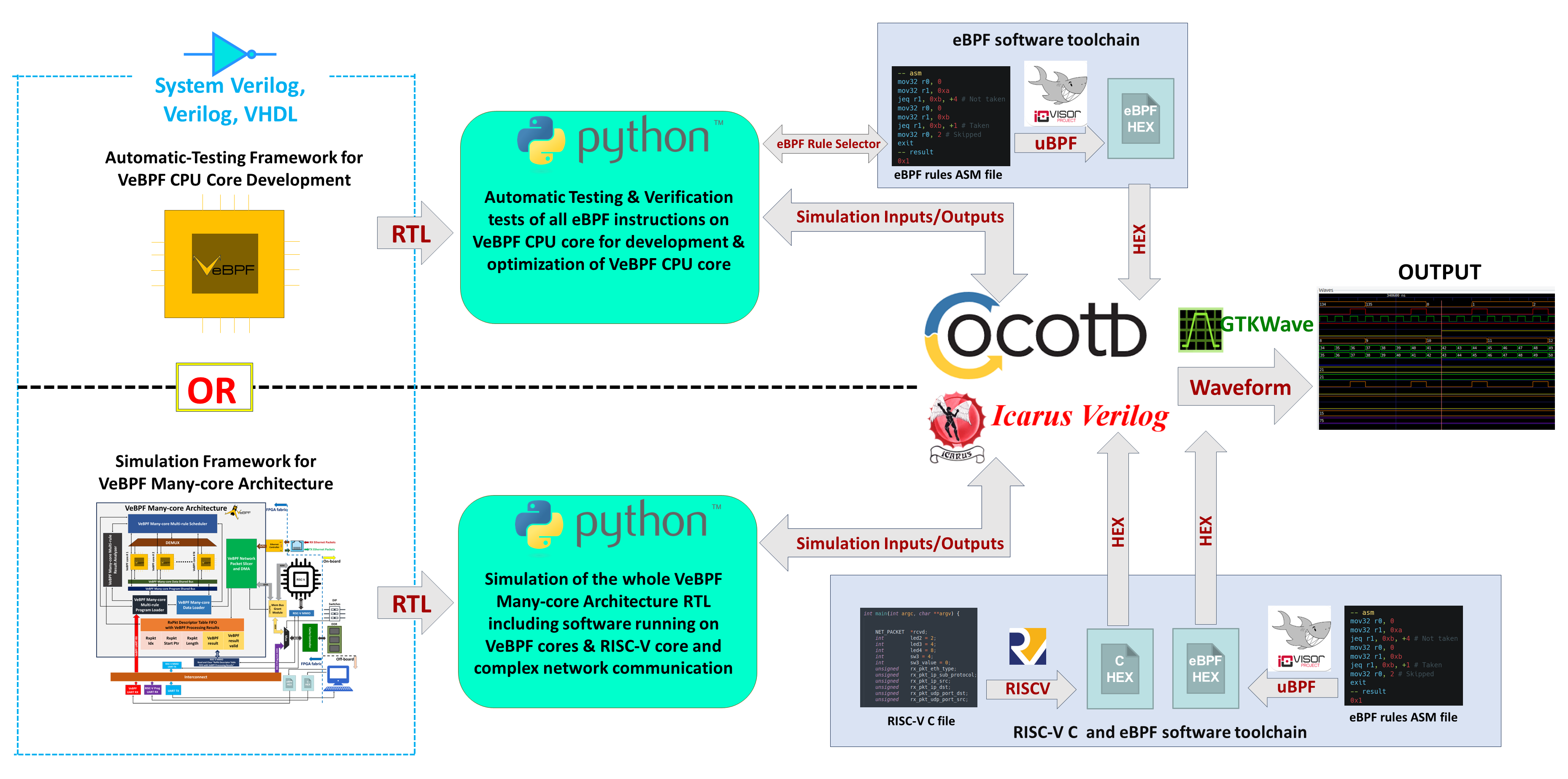}
\caption{VeBPF automatic-testing and simulation framework}
\label{fig8:VeBPF_sim}
\end{figure}

We have developed two separate automatic-testing and simulation frameworks (Fig.~\ref{fig8:VeBPF_sim}) for the further development and optimization of VeBPF CPU cores and the VeBPF many-core architecture respectively. Both automatic-testing and simulation frameworks have been developed using open-source tools like Python, Cocotb \cite{cocotb} and Icarus Verilog. Open-source tools like Python make it easy to simulate complex network packet interactions with our VeBPF many-core architecture. These automatic-testing and simulation tools are available on our publicly available code repositories and since they use open-source tools, anyone can use them. The VeBPF CPU core automatic-testing framework tests all the eBPF instructions on any upgrades made to the VeBPF CPU core and notifies about failures of the VeBPF core to execute any eBPF instruction, which streamlines and accelerates the development of VeBPF CPU core upgrades and the VeBPF-many core architecture.  


\section{Experiments and Results}
In order to show the high-configurability and versatility of the proposed VeBPF many-core architecture, we have implemented a firewall application against malicious network cyber attacks using the proposed VeBPF many-core architecture on a resource-limited FPGA-based IoT device. We used the FPGA board Arty A7-100T \cite{Arty} as the FPGA-based IoT device. For the eBPF firewall experiments, we chose to implement the state-of-the-art firewall rules being used by the technology sector as described by this technology report \cite{SANS}. Table-\ref{tab2:VeBPF_experiments} lists these firewall rules implemented for the experiments described below.

\begin{figure}[hbt!]
\begin{subfigure}{0.475\columnwidth} 
  \includegraphics[width=\columnwidth]
  {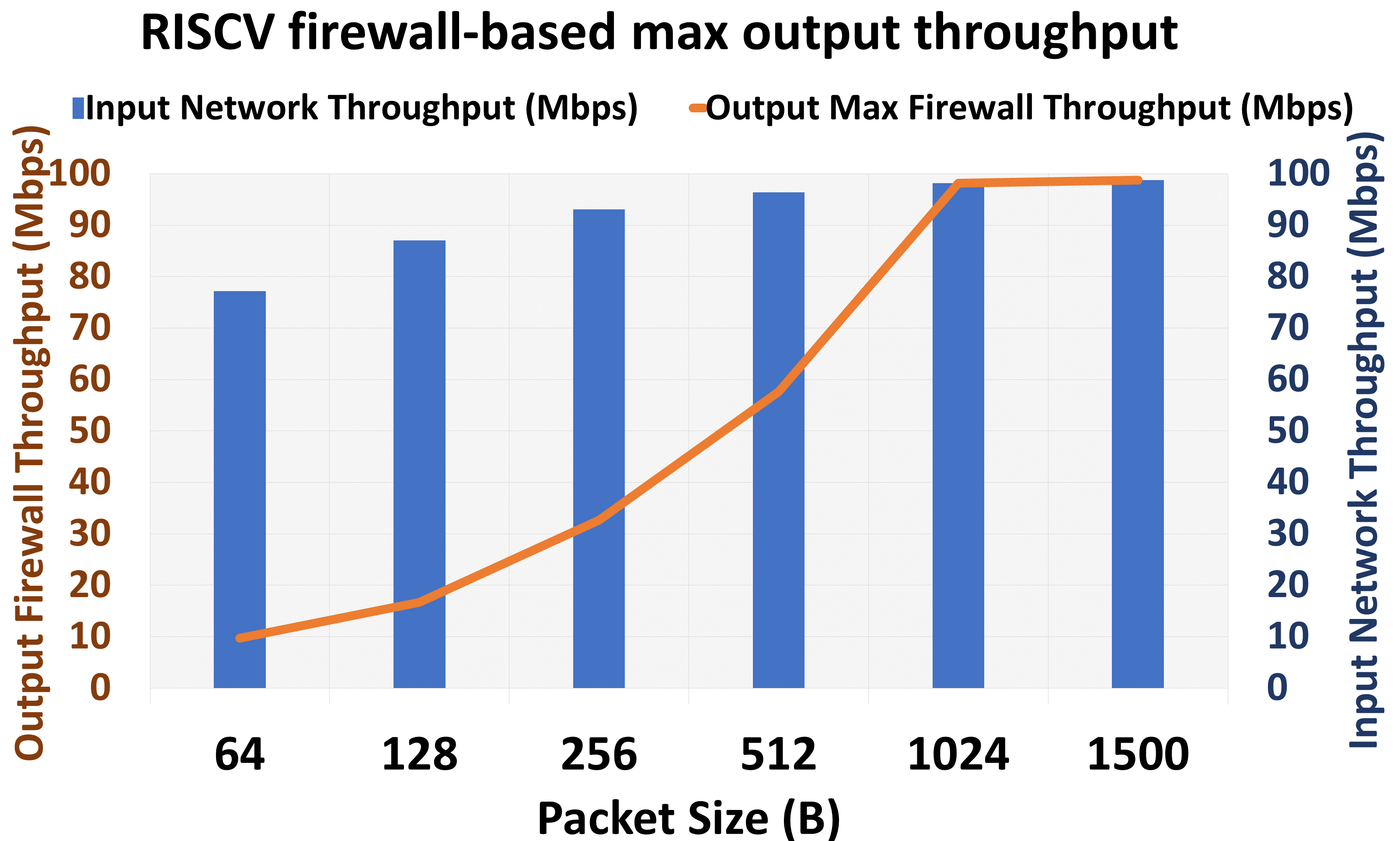}
\end{subfigure}\hfill 
\begin{subfigure}{0.475\columnwidth}
  \includegraphics[width=\columnwidth]
  {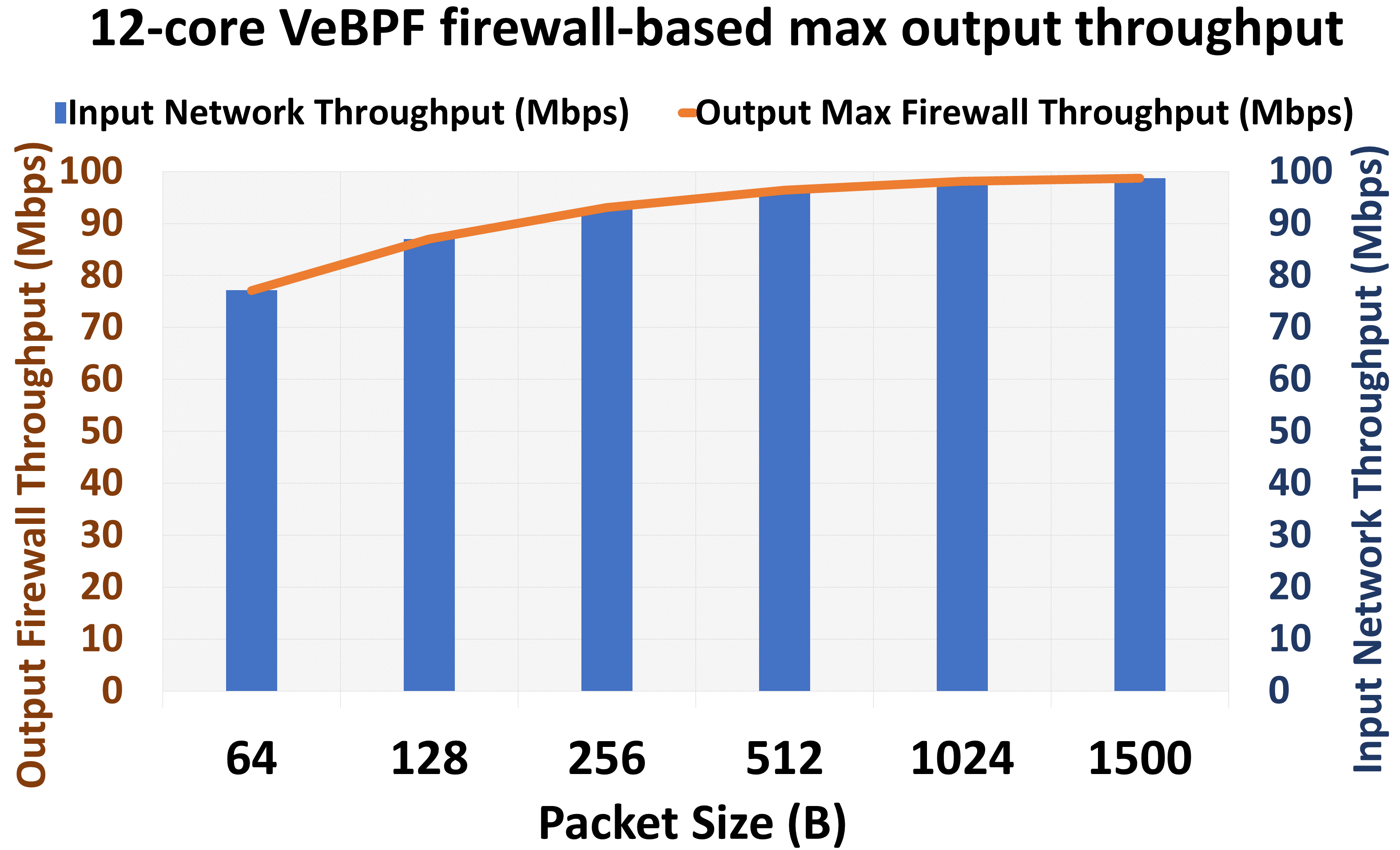}
\end{subfigure}
\begin{subfigure}{0.475\columnwidth}
  \includegraphics[width=\columnwidth]
  {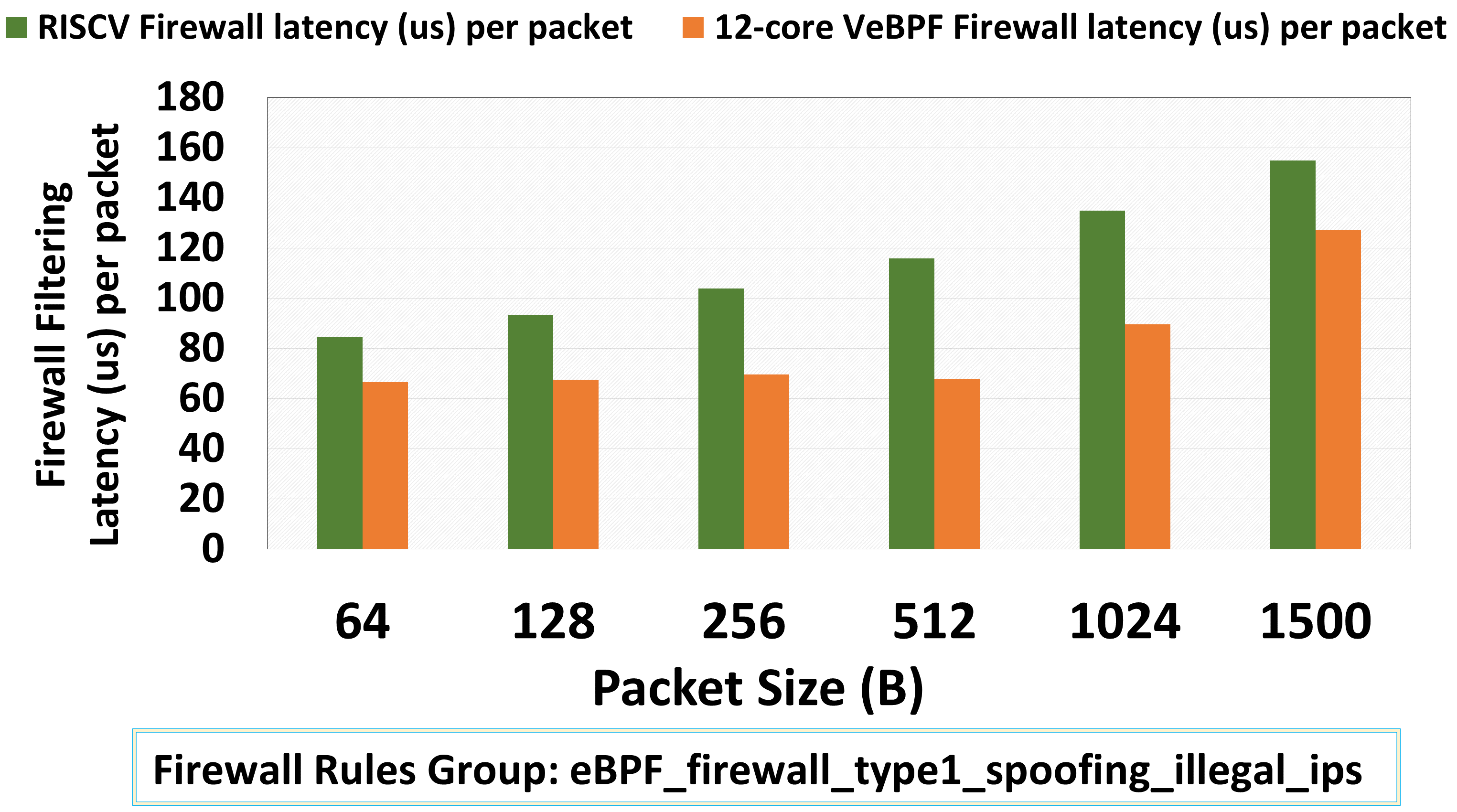}
\end{subfigure}\hfill 
\begin{subfigure}{0.475\columnwidth}
  \includegraphics[width=\columnwidth]
  {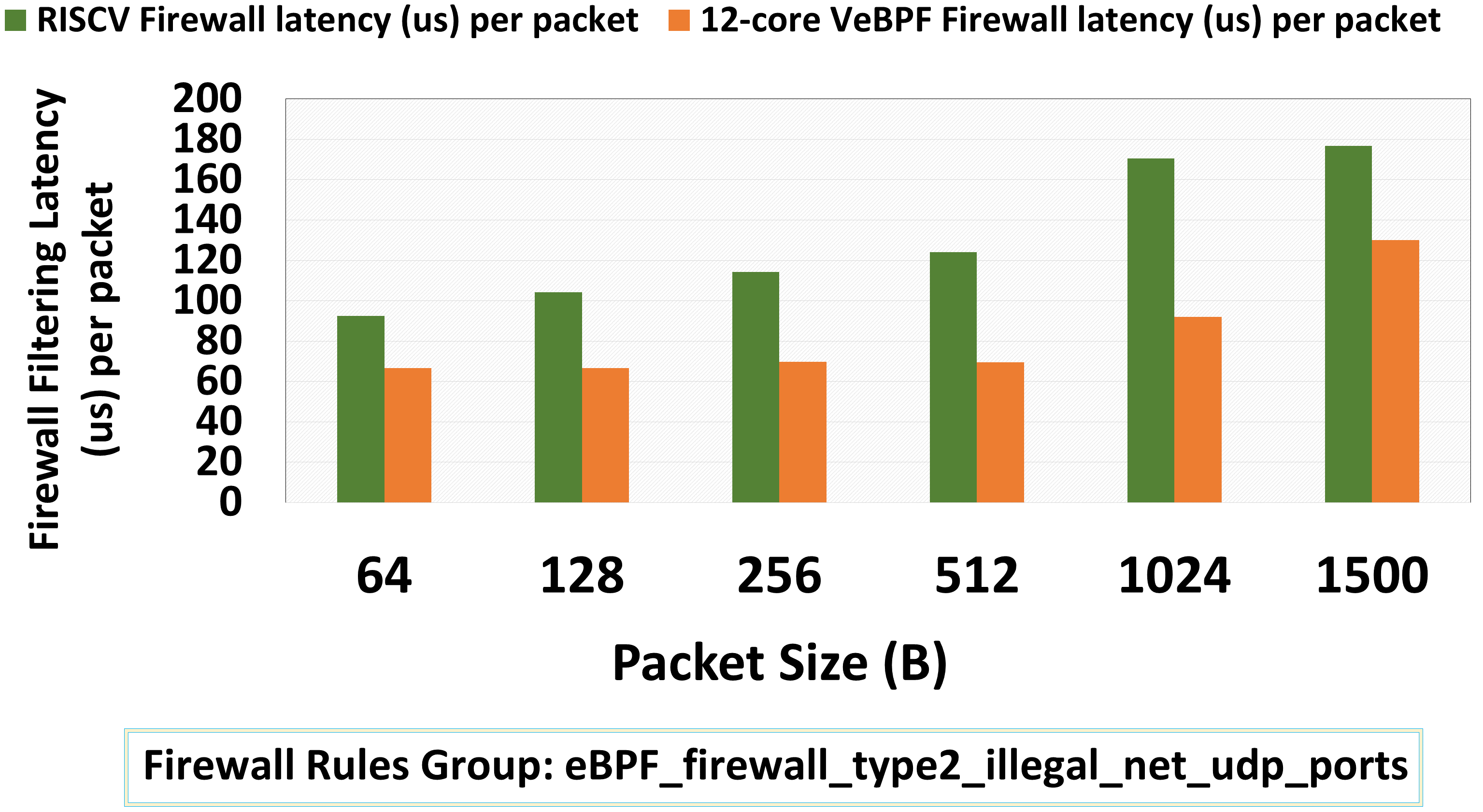}
\end{subfigure}
\begin{subfigure}{0.475\columnwidth}
  \includegraphics[width=\columnwidth]{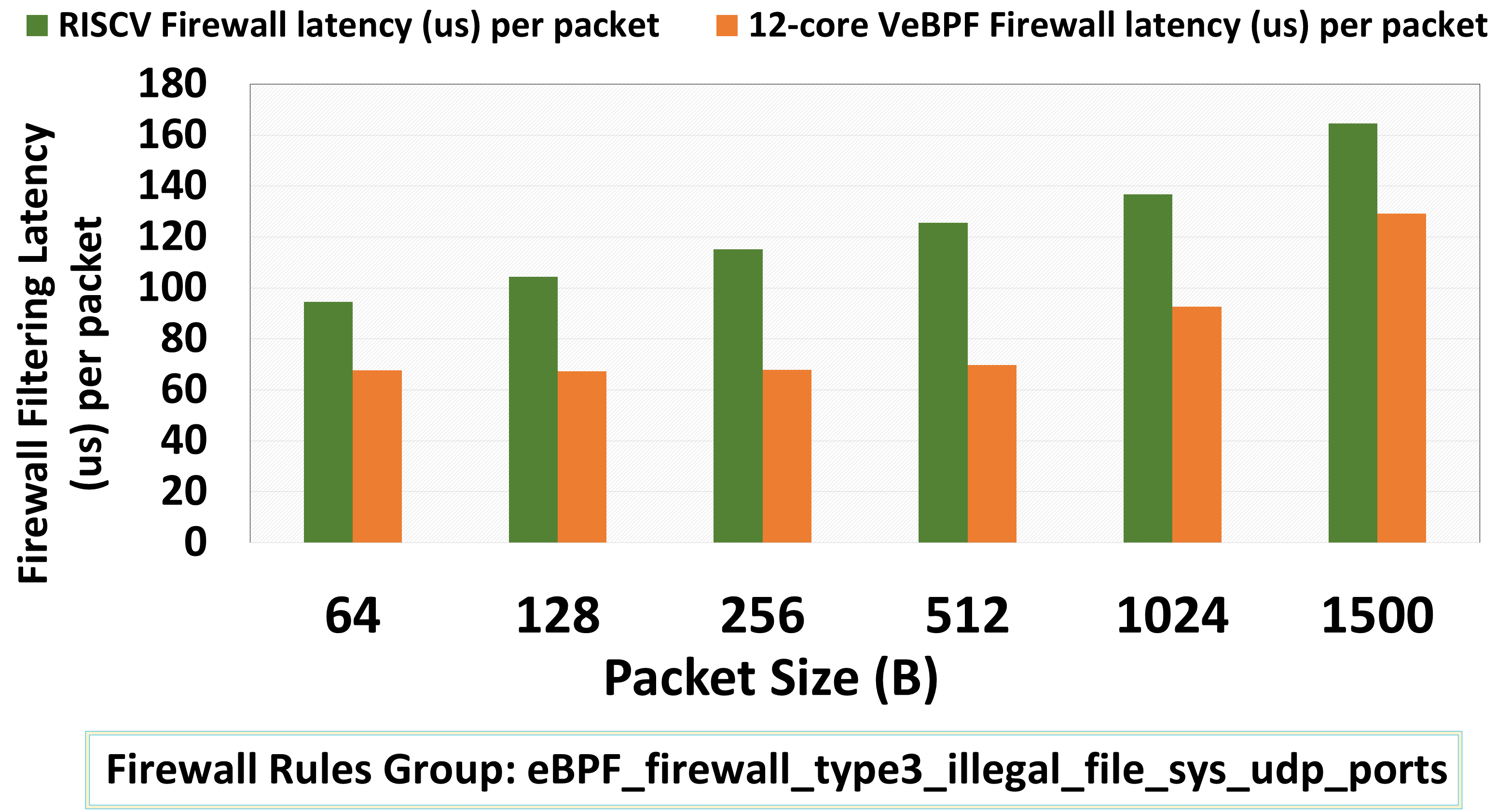}
\end{subfigure}
\begin{subfigure}{0.475\columnwidth}
  \includegraphics[width=\columnwidth]{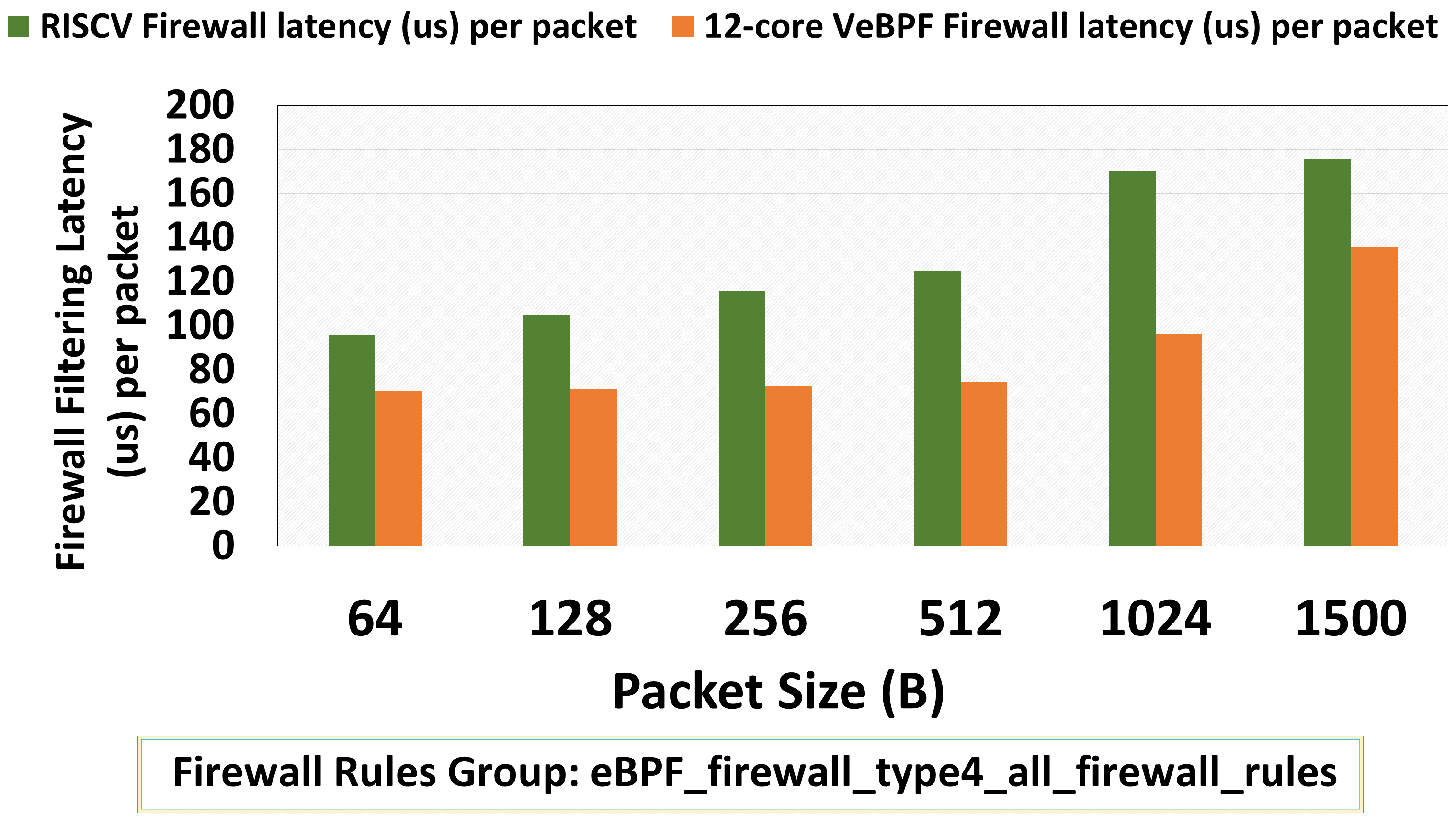}
\end{subfigure}

\caption{Firewall throughput and latency performance comparison between VeBPF many-core firewall and RISC-V.}
\label{fig5:VeBPF_experiments}
\end{figure}

For evaluation of our VeBPF many-core architecture firewall application (VeBPF many-core firewall). we uploaded the Table-\ref{tab2:VeBPF_experiments} firewall rules as eBPF bytecode with the total number of VeBPF cores \textit{N\textsub{VeBPF}} set at 12, since it was the highest number of VeBPF cores we could synthesize along with all the required subsystems as shown in Fig.~\ref{fig2:VeBPF_arch}, before the resources ran out on the FPGA board. We evaluated the performance of the 12-core VeBPF many-core firewall by comparing its performance (latency) versus the RISC-V performance for filtering the incoming network packets according to the Table-\ref{tab2:VeBPF_experiments} firewall rules. For the VeBPF many-core firewall evaluation we sent 2000 malicious \textit{rx\_pkts} conforming to each firewall rule in Table-\ref{tab2:VeBPF_experiments} from the host PC at 100 Mbps to the Arty FPGA board through the ethernet port. We repeated each experiment multiple times to cater for randomization. Also, all the experiments were repeated for different sizes of the \textit{rx\_pkts} as shown in Fig.~\ref{fig5:VeBPF_experiments}. 

The max output firewall throughput of the 12-core VeBPF many-core firewall versus RISC-V firewall can be seen in Fig.~\ref{fig5:VeBPF_experiments} (top two graphs). It is noted here that the 12-core VeBPF many-core firewall processes the \textit{rx\_pkts} at line-rate for all packet sizes as seen by the orange line (output firewall throughput) following the blue bars (input network throughput). Whereas the RISC-V soft-processor on the FPGA-based IoT device isn’t able to filter the smaller sized malicious \textit{rx\_pkts} at line-rate, even after being fully committed to filtering the network packets as per the firewall rules and not performing any other IoT functions as seen in Fig.~\ref{fig5:VeBPF_experiments} (top two graphs). 

Hence the two-fold advantage that our 12-core VeBPF many-core firewall provides is: 

1) Line-rate firewall filtering of incoming network packets;

2) The VeBPF firewall carries the firewall processing load, hence freeing up the RISC-V soft-processor to perform all the safety and mission critical IoT device related tasks;

The Fig.~\ref{fig5:VeBPF_experiments} (bottom 4 graphs) also displays the performance comparison in terms of firewall filtering latency per network packet for the 12-core VeBPF many-core firewall versus the RISC-V soft-processor for the firewall rules mentioned in Table-\ref{tab2:VeBPF_experiments}. We notice a common trend that the 12-core VeBPF many-core firewall outperforms the RISC-V soft-processor for all packet sizes even after the RISC-V soft-processor is fully committed to firewall processing. The RISC-V firewall filtering results would have been way worse if the RISC-V soft-processor was performing other IoT related tasks as well, like a real IoT device, which further highlights the advantages provided by the proposed VeBPF many-core architecture for network packet processing using native eBPF bytecode. 

\begin{table}[h]
    \tiny
    \centering
    \caption{Firewall Rules}
    \begin{tabular}{|c|c|} \hline 
 \textbf{Firewall Rule Types}& \textbf{Blocked IPs, UDP ports}\\ \hline  Type-1 (Illegal \& Spoofing source IPs)&  255.255.255.255, 127.0.0.0, 240.0.0.0, 0.0.0.0\\ \hline
 Type-2 (Network related critical destination UDP ports)&111, 2000, 37, 135, 137, 138, 161, 162, 514\\\hline
 Type-3 (File system related critical destination UDP ports)&69, 2049, 389, 4045\\\hline
 Type-4 (All rules combined)&All 3 firewall rule types combined\\\hline
    \end{tabular}
    \label{tab2:VeBPF_experiments}
\end{table}

\section{Conclusion and Future Work}
In this paper we presented a VeBPF many-core architecture that provides network packet processing functionality for both high-end and low-end FPGAs for FPGA target deployments like SmartNICs and IoT. This VeBPF many-core architecture is built using VeBPF CPU cores developed by us as the PE of this many-core architecture. These VeBPF cores are eBPF ISA compliant and are specialized for network packet processing and use native eBPF bytecode as program memory.

Our experimentation on the FPGA-based IoT device as the target deployment of the proposed VeBPF many-core architecture shows that the VeBPF many-core firewall implements state-of-the-art firewall rules for incoming malicious network attacks and filters them at line-rate and takes the processing load off the main IoT device processor (RISC-V) as compared to a dedicated RISC-V soft-processor which is slower.

For future work we want to show the proposed VeBPF many-core architecture implemented on a FPGA-based SmartNIC and HPC applications. We are also looking into adding more configuration options in the VeBPF many-core architecture. We have released the code, for VeBPF CPU core and VeBPF many-core architecture Verilog HDL along with their simulation frameworks built using open-source tools and the C libraries for the RSIC-V m-plane of the VeBPF many-core architecture, as an open-source contribution for further advancement of FPGAs in many-core architectures and communication.

\section*{Acknowledgments}

This work was supported, in part, by Red Hat through award 2024-01-RH08.

\bibliographystyle{IEEEtran}

\end{document}